\documentclass[sn-mathphys-num]{sn-jnl}

\usepackage{graphicx}%
\usepackage{multirow}%
\usepackage{amsmath,amssymb,amsfonts}%
\usepackage{amsthm}%
\usepackage{mathrsfs}%
\usepackage[title]{appendix}%
\usepackage{xcolor}%
\usepackage{textcomp}%
\usepackage{manyfoot}%
\usepackage{booktabs}%
\usepackage{algorithm}%
\usepackage{algorithmicx}%
\usepackage{algpseudocode}%
\usepackage{listings}%
\usepackage{subfigure}

\theoremstyle{thmstyleone}%
\theoremstyle{thmstyletwo}%

\theoremstyle{thmstylethree}%

\begin{document}

\title[title]{The Sliding Flux Ramp Demodulation Algorithm with High Sampling Rate in Microwave SQUID Multiplexer}


\author[1,2]{\fnm{Guofu}\sur{Liao}}

\author*[1]{\fnm{Congzhan}\sur{Liu}}\email{liucz@ihep.ac.cn}
\author*[1]{\fnm{Zhengwei}\sur{Li}}\email{lizw@ihep.ac.cn}
\author*[1]{\fnm{Daikang}\sur{Yan}}\email{yandk@ihep.ac.cn}
\author[3]{\fnm{Xiangxiang}\sur{Ren}}
\author[1]{\fnm{Yongjie}\sur{Zhang}}
\author[1]{\fnm{Laiyu}\sur{Zhang}}
\author[1,2]{\fnm{Yu}\sur{Xu}}
\author[1]{\fnm{Shibo}\sur{Shu}}
\author[1]{\fnm{He}\sur{Gao}}
\author[1]{\fnm{Yifei}\sur{Zhang}}
\author[1]{\fnm{Xuefeng}\sur{Lu}}
\author[1]{\fnm{Xufang}\sur{Li}}
\author[1]{\fnm{He}\sur{Xu}}
\author[1]{\fnm{Di}\sur{Wu}}

\affil[1]{\orgdiv{Institute of High Energy Physics}, \orgname{Chinese Academy of Sciences}, \orgaddress{\street{19B Yuquan Road}, \city{Shijingshan District}, \postcode{100049}, \state{Beijing}, \country{China}}}

\affil[2]{\orgdiv{University of Chinese Academy of Sciences}, \orgname{Chinese Academy of Sciences}, \postcode{100049}, \state{Beijing}, \country{China}}

\affil[3]{\orgdiv{Key Laboratory of Particle Physics and Particle Irradiation (MOE), Institute of Frontier and Interdisciplinary
Science, Shandong University}, \city{Qingdao}, \postcode{266237}, \state{Shandong}, \country{China}}



\abstract{
 Microwave SQUID Multiplexing ($\mu$MUX) is a widely used technique in the low-temperature detectors community as it offers a high capacity for reading large-scale Transition-Edge Sensor (TES) arrays. This paper proposes a Sliding Flux Ramp Demodulation (SFRD) algorithm for $\mu$MUX readout system. It can achieve a sampling rate in the order of MHz while maintaining a multiplexing ratio of about one thousand. Advancing of this large array readout technique makes it possible to observe scientific objects with improved time resolution and event count rate.  This will be highly helpful for TES calorimeters in X-ray applications, such as X-ray astrophysics missions.
}



\maketitle

\section{Introduction}\label{sec:intro}

The $\mu$MUX is invented to enable the readout of large arrays of TESs. As a superconducting thermal equilibrium detector, the latter has an excellent signal-to-noise ratio and shows great potential in science applications that require high energy resolution or low noise equivalent power (NEP). Compared with other TES multiplexing techniques, known as Time-Division Multiplexer, Code-Division Multiplexer, and Frequency-Division Multiplexer, $\mu$MUX requires fewer wires that run from the room-temperature stage to the cryogenic stage, which makes it attractive to TES involved experiments that need large-scale TES arrays or compact, low cooling-power detector chambers. 

One field that relies on large-scale TES arrays is cosmic microwave background (CMB) research. The demand for CMB studies involves scaling up to tens of thousands of pixels or more. The $\mu$MUX with a high multiplexing factor of about 1000 to 2000 has been used in several recently deployed and under-construction CMB telescopes, such as SO-SAT \cite{2024_galitzki_SimonsObservatoryDesignIntegrationTesting}
and AliCPT-1 \cite{2020_salatino_DesignAliCMBPolarizationTelescope}. 
Another scenario is space-based X-ray astrophysics telescopes with high-energy resolution. Several space-based telescopes for X-ray astrophysics have been proposed to use thousands of pixels from TES calorimeters as focal-plane detectors in recent years \cite{2023_bandler_LineEmissionMapperMicrocalorimeterSpectrometer,2023_barret_AthenaXrayIntegralFieldUnit,2021_gottardi_ReviewXrayMicrocalorimetersBasedSuperconducting,2020_sato_SuperDIOSMissionExploringDark,2020_cui_HUBSDedicatedHotCircumgalacticMedium,2019_bandler_LynxXrayMicrocalorimeter}. 
Driven by the need to measure finer energy spectra, structures, and temporal variability of astrophysical X-ray sources, large-scale TES arrays and high sampling rates are required. Additionally, a compact detector chamber is preferred in space. Meeting these requirements poses a significant challenge, as a high sampling rate necessitates a high flux ramp frequency, leading to high resonator bandwidth and frequency spacing requirements. Increasing the sampling rate often results in a decrease in the multiplexing ratio.

In this study, we propose a Sliding Flux Ramp Demodulation (SFRD) method to enhance the effective sampling rate of the $\mu$MUX system while maintaining a high multiplexing ratio.


\section{Algorithm}\label{sec:2}


\subsection{The $\mu$MUX Signal Modulation and Demodulation}\label{subsec:2-1}

With $\mu$MUX, the current signal from a TES is inductively coupled to an RF-SQUID (Radio Frequency Superconducting Quantum Interference Device) and then inductively coupled to a microwave resonator. By accommodating multiple such microwave resonators of different resonant frequencies into a common feedline, a number of TESs can be read out through one coaxial cable simultaneously. Each TES channel is measured by the S21 response of a microwave resonator:
\cite{2012_jonaszmuidzinas_SuperconductingMicroresonatorsPhysicsApplications}: 

\begin{equation}
    S_{21} = 1 - \frac{Q_r}{Q_c}\frac{1}{1+2j Q_r\frac{f_{exc}-f_r}{f_r}}
    \label{equ:s21}
\end{equation}

where $Q_r$ is the total quality factor and $Q_c$ is the coupling quality factor of the resonator, $f_{exc}$ is the fixed excitation frequency on the microwave feedline, and $f_r$ is the resonant frequency that changes with RF-SQUID inductance $L(\phi)$
\cite{2011_John_Mates_TheMicrowaveSQUIDMultiplexer}: 

\begin{align}
    f_r = \frac{f_1}{1+4f_1 C_c Z_1 + 4f_1 L(\phi) / Z_1} \label{equ:fr}, \\
    L(\phi) = L_c-\frac{M_c^2}{L_s}\frac{\lambda \cos(\phi)}{1+\lambda \cos(\phi)}, \label{equ:L_phi}\\
    \lambda \equiv \frac{L_S}{L_J}, \quad \phi \equiv 2\pi \frac{\Phi}{\Phi_0}.\label{equ:lambda_phi}
\end{align}

In Eq. (\ref{equ:fr}), $f_1$ is the intrinsic resonant frequency, $C_c$ is the coupling capacitance from the resonator to the common feedline, and $Z_1$ is the intrinsic impedance of the resonator.
In Eq. (\ref{equ:L_phi}), $L_c$ is the coupling inductance on the resonator to the RF-SQUID, $L_s$ is the self-inductance of the RF-SQUID that originates from its loop structure, and $M_c$ is the mutual inductance between the inductor and the RF-SQUID. In Eq. (\ref{equ:lambda_phi}), $\Phi$ is the magnetic flux in the RF-SQUID loop, and $\Phi_0$ is the magnetic flux quantum. 

By simply transforming Eq \ref{equ:s21}, we can obtain :

\begin{equation}
    S_{21} = 1 - \frac{Q_r}{2Q_c} - \frac{Q_r}{2Q_c} e^{-2j\arctan(2Q_r\frac{f_{exc}-f_r}{f_r})}
    \label{equ:s21_r}
\end{equation}

Here, when the excitation frequency $f_{exc}$ is constant, it can be seen that the trajectory of $S_{21}$ forms an arc of a circle on the I-Q plane. The center of the circle is  $1 - \frac{Q_r}{2Q_c}$ , the radius is  $\frac{Q_r}{2Q_c}$ , and the angle of the arc is related to $f_r$.

Assuming  $f_{exc} = \frac{f_1}{1+4f_1 C_c Z_1 + 4f_1 L_c / Z_1}$ , then we can get

\begin{equation}
    2Q_r \frac{f_{exc} - f_r}{f_r} = -2Q_r \frac{4f_1\frac{M_c^2}{L_s Z_1}\frac{\lambda\cos \phi}{1+\lambda \cos \phi}}{1+4f_1 C_c Z_1+4f_1 L_c / Z_1}.
    \label{equ:x}
\end{equation}

Here, considering $1+4f_1 C_c Z_1+4f_1 L_c / Z_1 \approx 1$, and assuming $8 Q_r f_1\frac{M_c^2}{L_s Z_1} \ll 1$. Therefore,

\begin{equation}
    \arctan(2Q_r \frac{f_{exc} - f_r}{f_r}) \approx 2Q_r\frac{f_{exc} - f_r}{f_r} 
    = -k \frac{\lambda\cos \phi}{1+\lambda \cos \phi}
\end{equation}

here, $ 0 < k = \frac{8 Q_r f_1\frac{M_c^2}{L_s Z_1}}{1+4f_1 C_c Z_1+4f_1 L_c / Z_1} \ll 1$.
Considering a small value of $\lambda < 0.5$ \cite{2011_John_Mates_TheMicrowaveSQUIDMultiplexer, 2012_mates_FluxRampModulationSQUIDMultiplexing}, then Eq. \ref{equ:s21_r} can be further simplified to

\begin{equation}
    S_{21} \approx 1 - \frac{Q_r}{2Q_c} - \frac{Q_r}{2Q_c} e^{2j k\lambda \cos \phi}
    \label{equ:s21_j}
\end{equation}

Therefore, when we obtain the $S_{21}$ trajectory at the excitation frequency, we can remove the center of the circle by translation. Then, we use the arctangent to obtain $2k\lambda \cos \phi$, which is exactly the SQUID signal. In practice, the measured $S_{21}$ contains cable delay components, which will cause the $S_{21}$ arc trajectory to rotate by a certain angle on the complex plane. So, the practical process includes translation to move the center to zero (or the origin) and rotation to remove the cable delay before performing arctangent to obtain the SQUID signal $2k\lambda \cos \phi$.

The $\mu$MUX system adopts a sawtooth signal to linearly ramp the magnetic flux in the RF-SQUID, so called “flux-ramp modulation” \cite{2008_mates_DemonstrationMultiplexerDissipationlessSuperconductingQuantum}. Based on Eq. (\ref{equ:s21_j}), the flux ramp modulation leads to a sinusoidal response in the finally measured SQUID signal. The TES current signal, superimposed to the flux-ramp signal, will cause an extra phase shift in the sinusoidal SQUID response. Therefore, the measured signal has the form: 

\begin{equation}
\theta(t) = 2k \lambda\cos(2\pi f_{ramp} N_{\Phi_0}t + \phi_{TES}(t))
\label{equ:squid_t}
\end{equation}

where $f_{ramp}$ is the frequency of the sawtooth flux-ramp signal. $N_{\Phi_0}$, usually an integer, is the number of $\Phi_0$ the ramp signal generates in the RF-SQUID, and is directly controlled by the ramp signal amplitude. 
$\phi_{TES}(t)=2\pi\frac{\Phi_{TES}(t)}{\Phi_0}$ represents the phase shift caused by the TES signal. $\Phi_{TES}$ represents the magnetic flux caused by the TES signal in the SQUID loop.

According to the Eq \ref{equ:squid_t}, it can be seen that the phase shift of the SQUID signal is proportional to the TES signal.
When $2\pi f_{ramp} N_{\Phi_0}$ is much greater than the slope of $\phi_{TES}(t)$,
this phase shift can be solved as

\begin{equation}
    \phi_{TES} = \arctan(-\frac{\int \theta(t)\sin(\omega_{mod}t)}{\int \theta(t) \cos(\omega_{mod}t)})
    \label{equ:frd_t}
\end{equation}

Here, $\omega_{mod}=2\pi f_{ramp} N_{\Phi_0}$ represents the modulation angular frequency.
The integration interval spans an integer number of modulation cycles.
This process is referred to as Flux Ramp Demodulation (FRD).

Considering the discrete nature of the digitizing system, with a sampling frequency of $f_s$, the actual demodulated TES signal is: 

\begin{equation}
\phi_{TES}'[m] = \arctan\big(-\frac{\sum_{n=Nm}^{N(m+1)-1}\theta[n]\sin(\Omega_{mod}n)}{\sum_{n=Nm}^{N(m+1)-1}\theta[n]\cos(\Omega_{mod}n)}\big)
\label{equ:frd_n}
\end{equation}

Here, $\Omega_{mod} = 2\pi\frac{f_{ramp}N_{\Phi_0}}{f_s}$
represents the digital modulation angular frequency and $N=f_s/f_{ramp}$ is the number of data samples in one flux-ramp cycle.
The effective sampling rate of the $\mu$MUX system, therefore, is the flux-ramp frequency $f_{ramp}$.


\subsection{Sliding Flux Ramp Demodulation}\label{subsec:2-2}

Performing a Discrete Fourier Transform (DFT) to the time series $\theta[n]$ allows to extract $\phi_{TES}[n]$ as it will correspond to the phase angle of the Fourier coefficient of the fundamental component, of frequency $\Omega_{mod}$.
Since only the Fourier coefficient of the fundamental frequency is required, the calculation process can be simplified using the Sliding Discrete Fourier Transform (SDFT)\cite{2022_chauhan_RecursiveSlidingDFTAlgorithmsReview,2003_lyons_DspTipsTricksSlidingDFT}. When SDFT is applied, the update rate of $\phi_{TES}[n]$ is consistent with that of $\theta[n]$.
The recursive form of SDFT is shown as follows,

\begin{align}
    \alpha_i[n]&=(\alpha_i[n-1]+\theta[n]-\theta[n-M])e^{j2\pi \frac{i}{M}}; \quad i = 0,1,...,M-1
    \label{equ:sdft_a} \\
    \phi_{TES}[n] &= \angle \alpha_1[n]
    \label{equ:sdft_phi}
\end{align}

Here, $\alpha_i[n]$ represents the discrete Fourier coefficient of the $i$-th frequency at the $n$-th time point.
$M = \frac{2\pi}{\Omega_{mod}}$ denotes the period of digital modulation.
By definition of $\Omega_{mod}$, $M$ is also the number of sample points within one $\Phi_0$ period.

\begin{figure}[h]
\centering
\includegraphics[width=1\textwidth]{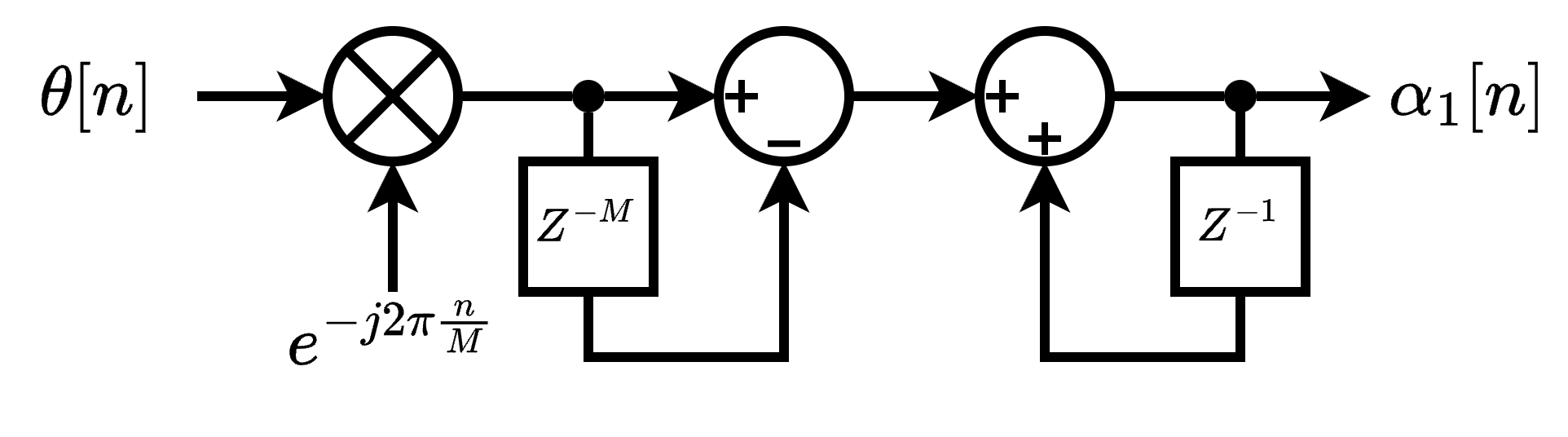}
\caption{The calculation structure of mSDFT. }
\label{fig:sfrd}
\end{figure}

Using this calculation method, the sampling rate of $\phi_{TES}$ is consistent with that of $\theta$, which is $f_s$. Due to the error accumulation and instability issues inherent in the standard Sliding Discrete Fourier Transform calculation structure, we adopt the Modulated Sliding Discrete Fourier Transform (mSDFT) here
\cite{2022_chauhan_RecursiveSlidingDFTAlgorithmsReview}. The mSDFT modulates the SQUID signal to the DC component (the 0-th frequency point) using a modulation sequence, and its computational structure is as follows (shown in Fig \ref{fig:sfrd}):

\begin{align}
    \theta_m[n] &= \theta[n] e^{-j2\pi \frac{n}{M}}
    \label{equ: sdft_theta} \\
    \alpha_1[n] &=\alpha_1[n-1]+\theta_m[n]-\theta_m[n-M]
    \label{equ: sdft_ma} \\
     \phi_{TES}[n] &=\angle\alpha_1[n]
     \label{equ:sdft_phi_1}
\end{align}

The mSDFT has the advantages of stability, good noise performance, and low error \cite{2022_chauhan_RecursiveSlidingDFTAlgorithmsReview}.
This is our sliding flux ramp demodulation (SFRD).
By using SFRD, the sampling rate of the TES signal is improved by a factor of $f_s/f_{ramp}$ compared to the standard flux ramp demodulation method. In general, $f_s/f_{ramp}$ will be a certain big number.


\section{Experiment and preliminary results}\label{sec:3}

We conducted preliminary hardware experiments to demonstrate the SFRD algorithm. We applied a $\mu$MUX modulation signal to an excitation frequency to simulate the excitation signal processed by the $\mu$MUX device. This signal was generated by a pair of Digital-to-Analog Converters (DACs) and collected by a pair of Analog-to-Digital Converters (ADCs). The collected data was channeled within the Field Programmable Gate Array (FPGA) and transmitted to the computer via Ethernet. On the computer, translation, rotation, arctangent operations, standard FRD and SFRD  were performed.

Table \ref{tbl:para} shows the parameter values chosen for our $\mu$MUX modulation signal model.
$\phi$ is the input to the $\mu$MUX modulation model, composed of the Flux ramp signal and the TES signal. Here, the flux ramp frequency $f_{ramp}$ is set to 100 kHz, and $N_{\Phi_0}=2$. 

\begin{table}[!h]
        \caption{The parameter values of our modulation signal model} 
        \begin{tabular}{ll}  
            \toprule
            parameters & value \\ \midrule
            $L_c$ & 77pH \\
            $M_c$ & 4pH \\
            $L_s$ & 50.1pH \\
            $C_c$ & 1.49fF  \\
             $\lambda$ & 0.42 \\
            $f_1$ & 5.27GHz \\
            $Z_1$ & $50 \Omega$ \\
            $Q_r$ & 8084 \\
            $Q_c$ & 25499 \\
        \bottomrule
        \end{tabular} 
        \label{tbl:para} 
\end{table}


\subsection{Experimental Platform}\label{subsec:3-1}

\begin{figure}[h]
\centering
\includegraphics[width=0.8\textwidth]{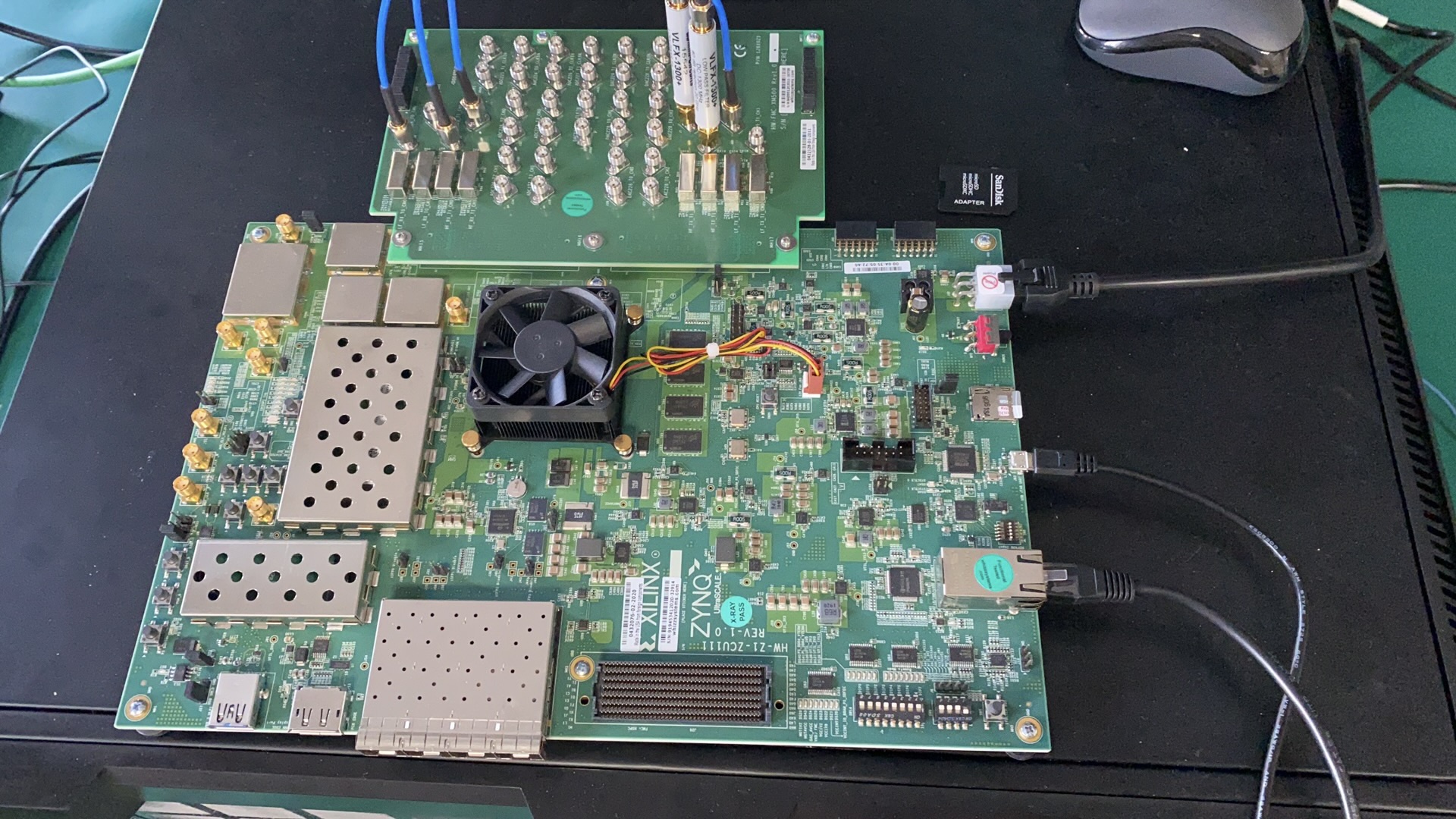}
\caption{
This is our experimental platform.
It consists of the Zynq UltraScale+ RFSoC ZCU111 evaluation board
and the XM500 RFMC Balun Add-on Card.
}
\label{fig:zcu111}
\end{figure}

\begin{figure}[h]
\centering
\includegraphics[width=0.7\textwidth]{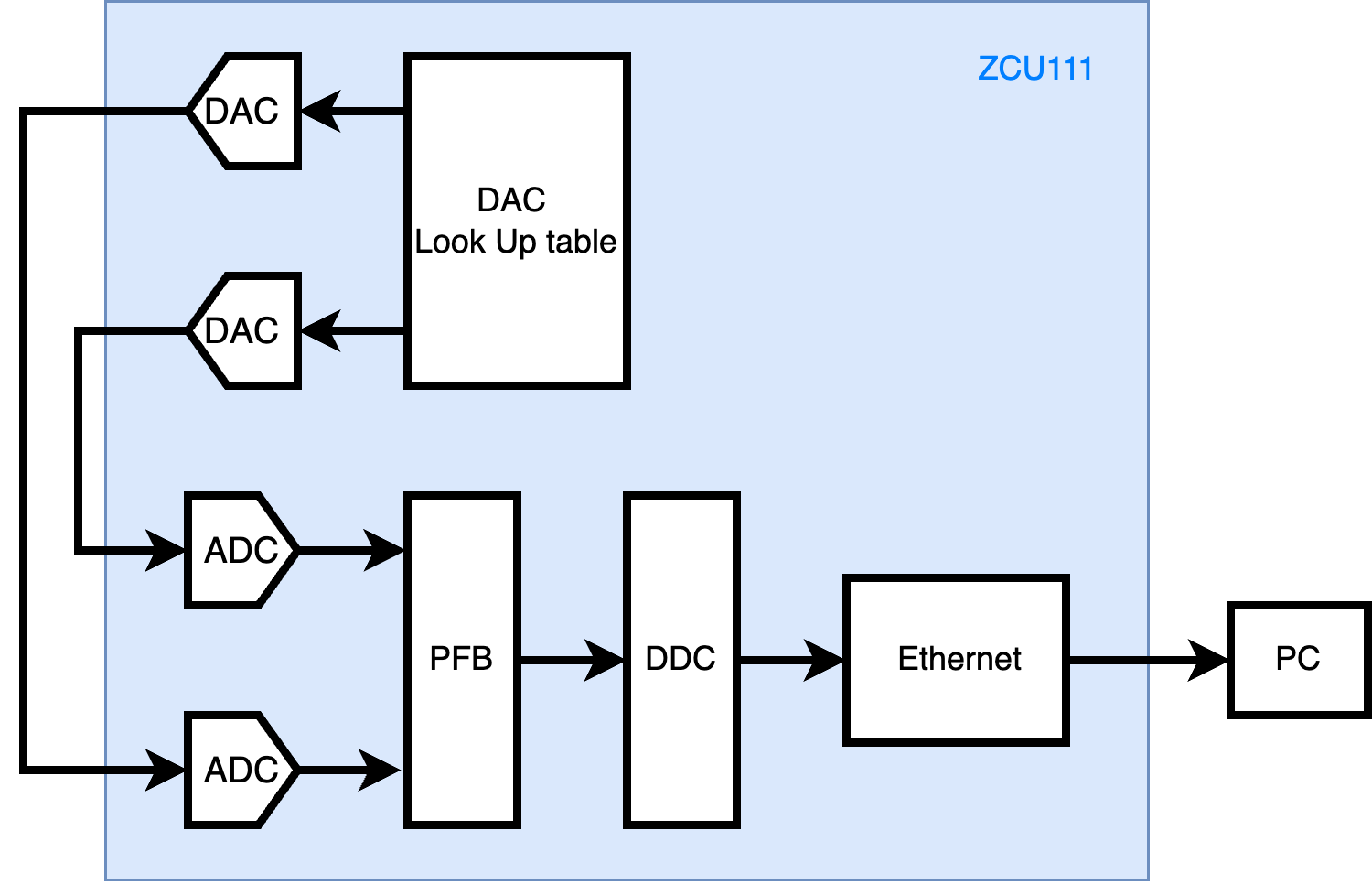}
\caption{The test diagram}
\label{fig:test_diagram}
\end{figure}

We utilize the Zynq UltraScale+ RFSoC ZCU111 evaluation board as our experimental platform, as shown in Fig \ref{fig:zcu111}. Both the ADC and DAC have a sampling rate set to 512 MHz. The excitation signal frequency used for modulation is 10 MHz. Fig \ref{fig:test_diagram} shows the test setup.  The “DAC Look Up Table” stores the excitation tones carrying the uMUX modulation signal in IQ format. The DAC converts these data into analog signals and then digitizes them by the ADC. Afterward, the 128-point polyphase filter bank (PFB) and digital down-conversion (DDC) are applied for channelization. 
So, the sampling rate for each channel decrease by a factor of 128 from 512MHz, that is 4MHz. After channelization, the selected channel is then transmitted via Ethernet to a PC for further data processing.

\subsection{X-ray pulse}\label{subsec:3-2}

\begin{figure}[h]
\centering
\includegraphics[width=0.9\textwidth]{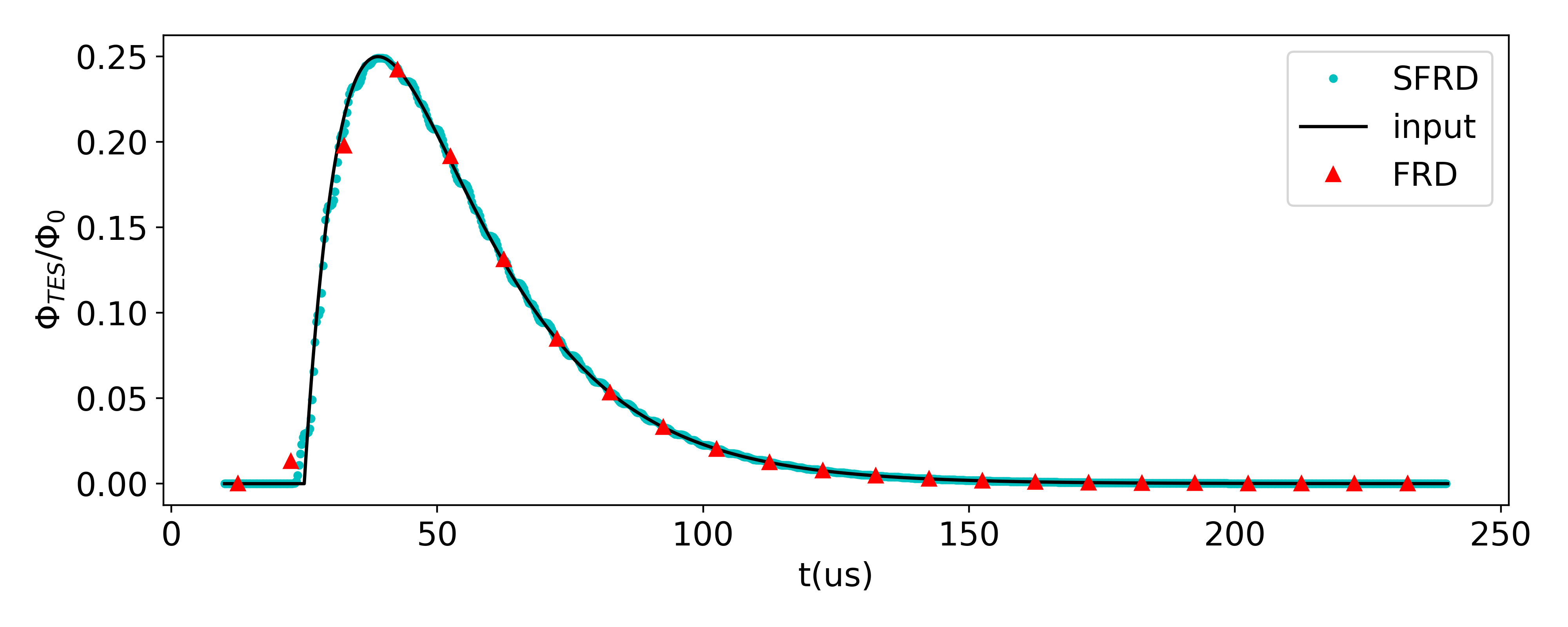}
\caption{
Comparison of Pulse Reconstruction between FRD and SFRD.
The 'input' is the TES signal input to the modulation model. 
The red triangles represent the TES signal solved by standard FRD, 
while the cyan dots represent the TES signal solved by the SFRD algorithm.
}
\label{fig:pulse}
\end{figure}

In X-ray astrophysics, the pulse shape of a TES detector can be generally described by a double exponential function. The detector’s electric and thermal time constants determine the pulse’s rising and falling times.
Here, we simulate a real TES pulse signal with a rising time constant $\tau_r = 10 \mu s$ and a falling time constant $\tau_f = 20 \mu s$. Compared to FRD, the sampling rate of SFRD is increased by a factor of $f_s/f_{ramp}=40$.
From Fig \ref{fig:pulse} , it can be observed that the reconstructed pulse signal by FRD consists of a few points, while SFRD can reconstruct the pulse waveform with finer time steps. This fully demonstrates the superior time resolution of SFRD. However, the non-zero slope of the rising and falling edges of the fast pulse signal leads to a deviation of the actual frequency of this part of the SQUID signal from the fundamental frequency. 
In this case, periodic ripples occur when calculating the phase at the fundamental frequency using a recursive structure. The amplitude of the ripples is positively correlated with the signal’s slope. 
These ripples can be observed at both the rising and falling edges in the Fig \ref{fig:pulse}. Fortunately, since the slope at the pulse peak is zero, no ripples occur at the pulse peak, making the calculation of the pulse height accurate. In Sec \ref{ssec:4-1} , we provide a detailed explanation of the generation of this slope-dependent ripple.

\subsection{Noise}\label{subsec:3-3}

We set the TES signal to be white noise to compare the noise performance of SFRD and FRD. Here, a total of 163440 points of SQUID signal were sampled, resulting in 4086 points obtained by FRD demodulation and 163420 points obtained by SFRD demodulation. 
The noise spectrum of SFRD is almost the same as FRD below the frequency of $f_{ramp}/2$. The noise is still kept on a certain low level above $f_{ramp}/2$.

\begin{figure}[h]
\centering
\includegraphics[width=0.9\textwidth]{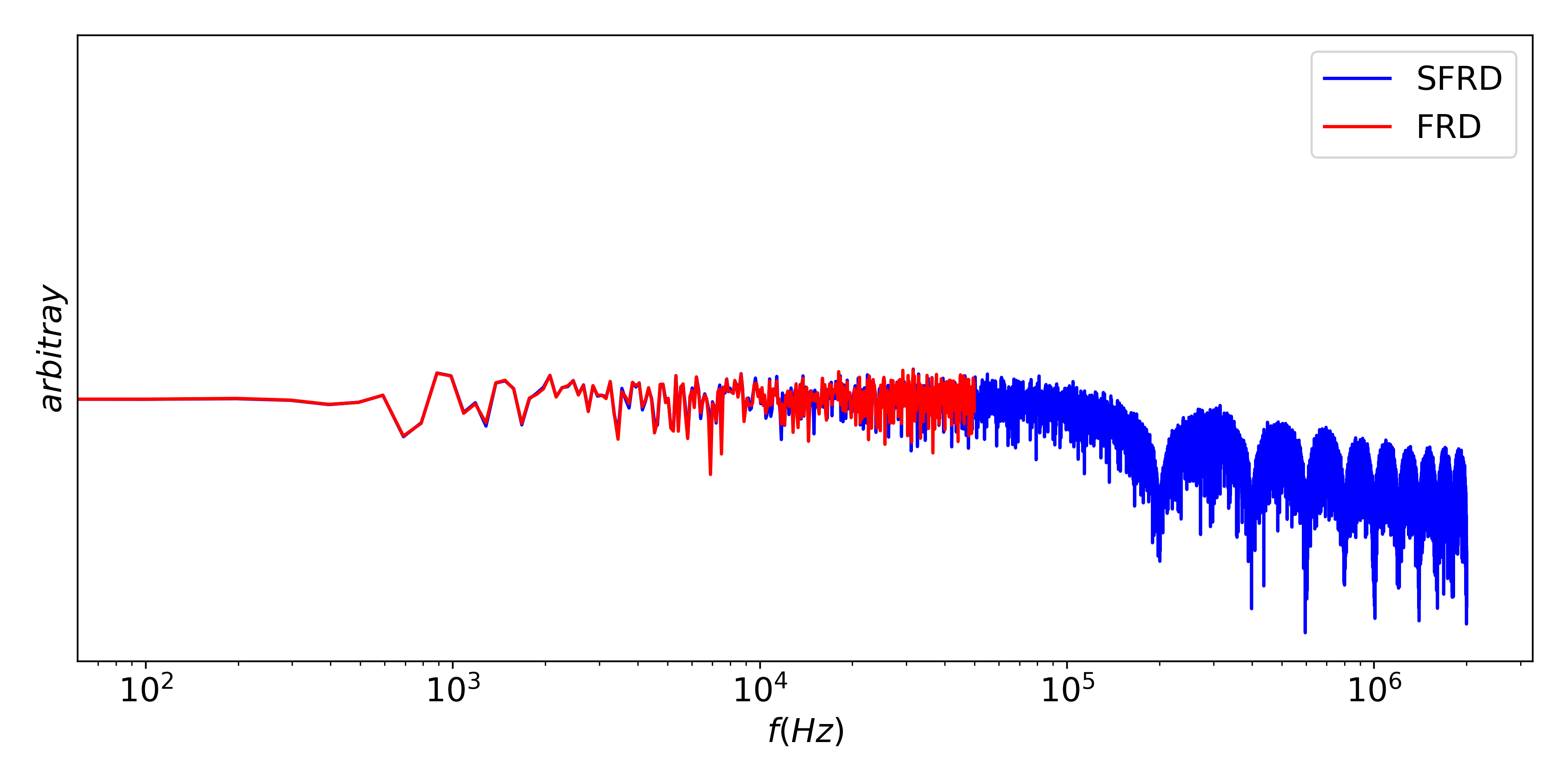}
\caption{
Comparison of Noise Performance between FRD and SFRD.
Since $f_{ramp} = 100$ kHz and $f_s = 4$ MHz,
the noise bandwidth of FRD is $f_{ramp}/2 = 50$ kHz,
and the noise bandwidth of SFRD is $f_s/2 = 2$ MHz.
}
\label{fig:noise_pdf}
\end{figure}

\subsection{Energy resolution}

\begin{figure}[h]
\centering
\includegraphics[width=1\textwidth]{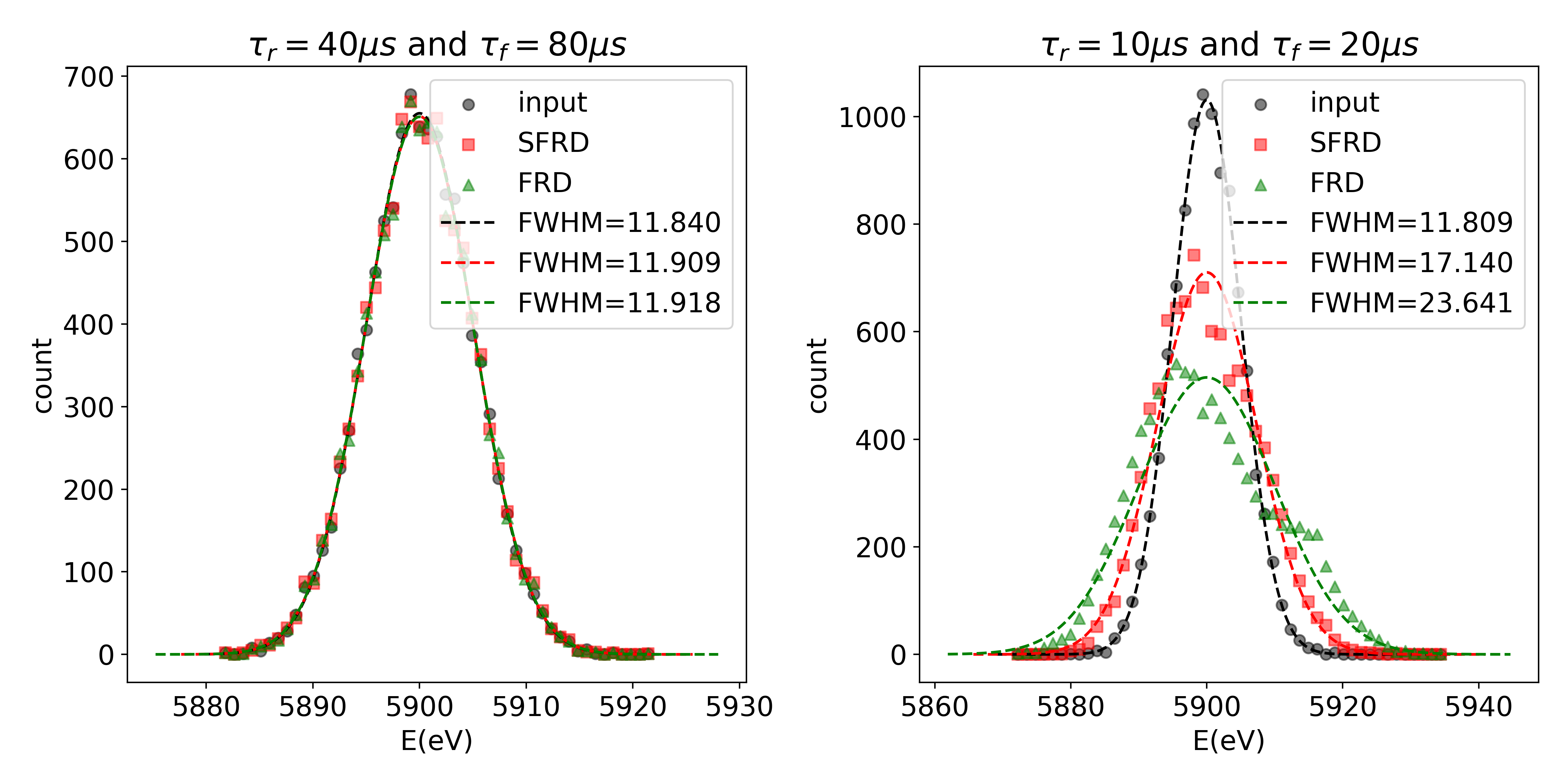}
\caption{
Comparison of energy spectra performance between FRD and SFRD.
The black dots represent the energy spectrum input into the modulation model, the red squares represent the energy spectrum obtained using the SFRD method, and the green triangles represent the energy spectrum obtained using the FRD method. The black, red, and green dashed lines are the Gaussian fits to the three statistical data sets. The left plot shows the results for slow pulses, and the right plot shows the results for fast pulses.
}
\label{fig:energy}
\end{figure}

Energy resolution is a crucial parameter in X-ray measurements. To evaluate the influence of the SFRD method on energy resolution, we compare the energy spectra obtained from SFRD with those obtained from standard FRD. We set up a simulation experiment where we assume that the pulse arrival times are uniformly randomly distributed within one flux ramp period, and the energies corresponding to the pulse amplitudes follow a Gaussian distribution with 11.8 eV@5.9 keV. We sample the pulse arrival times and pulse amplitudes 10000 times. We use SFRD and FRD to calculate the pulse data and create energy spectra. We consider both fast pulses ($\tau_r = 10 \mu s$ and $\tau_f = 20 \mu s$) and slow pulses ($\tau_r = 40 \mu s$ and $\tau_f = 80 \mu s$). Fig \ref{fig:energy} shows the resulting energy spectra.
The results show that the energy resolution of both SFRD and FRD for slow pulses can match the input energy resolution. However, for fast pulses,8the energy resolution of FRD deteriorates more significantly, leading to a deviation from a Gaussian shape in the spectral profile. In comparison, the energy resolution of SFRD is somewhat better.

\section{Discussion}\label{sec:4}

\subsection{Slope-Dependent Ripples}\label{ssec:4-1}

In the standard FRD algorithm, the TES signal is typically assumed to be constant within the demodulation window. However, this assumption is not accurate in the case of large slope of a real signal. A more general assumption is that the TES signal is linear within the demodulation window, meaning $\phi_{TES}(t) = bt + \phi_0$, where b is the slope, and $\phi_0$ is the segment’s starting point.

At this point, Eq \ref{equ:squid_t} becomes:

\begin{equation}
    \theta(t) = 2k\lambda\cos(\omega_{mod}t + bt + \phi_0)
\end{equation}

And Eq \ref{equ:frd_t} becomes:

\begin{equation}
    \phi_{TES} = \arctan\left( -\frac{\int_0^{\frac{2\pi}{\omega_{mod}}} \theta(t)\sin(\omega_{mod}t) dt}{\int_0^{\frac{2\pi}{\omega_{mod}}} \theta(t)\cos(\omega_{mod}t) dt} \right) = \arctan\left( \frac{1}{1 + \frac{b}{\omega_{mod}}} \tan(\phi_{TES,aver} ) \right)
\end{equation}

\begin{figure}[h]
\centering
\includegraphics[width=0.8\textwidth]{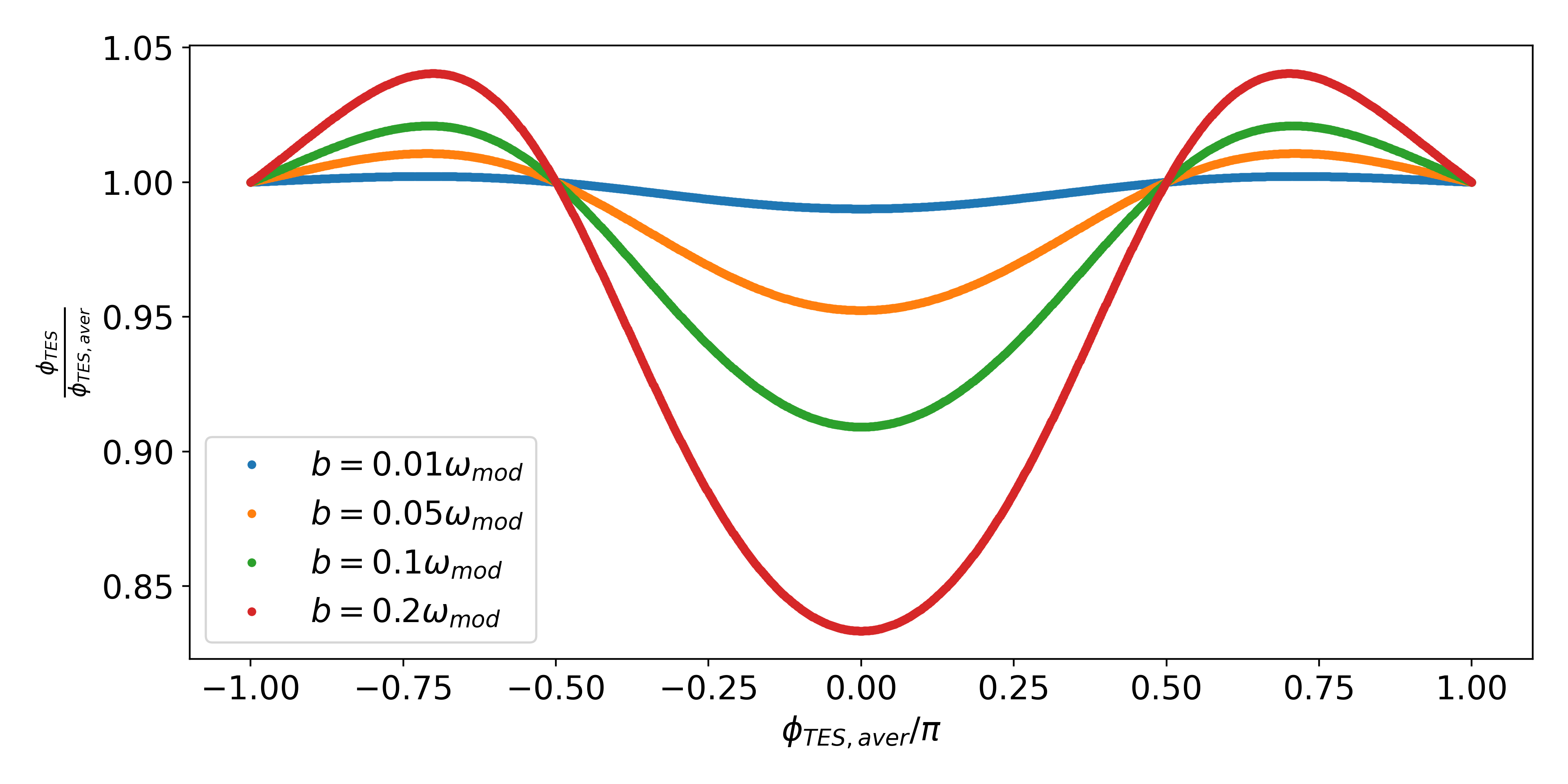}
\caption{The ratio between the calculated value and the true value under different slopes.}
\label{fig:ripple}
\end{figure}

Here, $\phi_{TES,aver}=\frac{b}{\omega_{mod}} \pi + \phi_0$ represents the segment’s average value, reflecting the true value, and $\phi_{TES}$ is the calculated value. 
We compared the ratio between the calculated and true values under different slopes, where the true value ranges from $-\pi$ to $\pi$ (Fig \ref{fig:ripple}). 
When the slope is fixed, the difference between the calculated and true values fluctuates as the true value changes. In our SFRD algorithm, since the demodulation window slides over time, the true value changes, leading to ripples in the calculated result. Additionally, as shown in Fig \ref{fig:ripple}, the larger the slope, the greater the ripple effect.

\section{Conclusion}\label{sec:5}

We present a study demonstrating the effectiveness of the Sliding Flux Ramp Demodulation (SFRD) algorithm in accurately tracking fast TES detector signals while maintaining low noise levels and achieving superior energy resolution. 
This work proposes the SFRD algorithm to overcome the trade-off between sampling rates and multiplexing ratios in $\mu$MUX systems.
It makes the implementation of a multiplexing ratio of over a thousand and sampling rates up to MHz feasible in $\mu$MUX systems.
Based on the experimentation on our prototype platform, the SFRD algorithm successfully achieved a sampling rate of 4 MHz (equivalent to the channelization rate), which can effectively reconstruct pulse signals with a rising time constant of 10 $\mu$s. Compared to the conventional FRD method, the SFRD algorithm demonstrates a significant advantage in high sampling rates without compromising noise levels below the $f_{ramp}/2$ frequency. However, the increased bandwidth resulting from the higher sampling rate introduces high-frequency noise components, contributing to an overall elevation in the noise level. Furthermore, our study reveals that the SFRD algorithm outperforms FRD regarding energy resolution for fast X-ray TES pulses.

We have implemented our SFRD algorithm on FPGA and plan to use $\mu$MUX devices to test our SFRD algorithm.


\bmhead{Acknowledgments}

This work is supported by the funds of the National Key Research and Development Program of China (Grant No. 2022YFC2204900, No. 2021YFC2203400), the Scientific Instrument Developing Project of the Chinese Academy of Sciences(Grant No.YJKYYQ20190065).

\bmhead{Author Contributions}

The concept of the paper originated from Guofu Liao. The overall framework of the paper was refined by Congzhan Liu, Zhengwei Li, Daikang Yan, and Xiangxiang Ren. Guofu Liao and Xiangxiang Ren conducted all the experiments. Guofu Liao, Congzhan Liu, and Daikang Yan completed the main manuscript, and Zhengwei Li participated in the revision of the paper. All authors reviewed the manuscript.

\bmhead{Availability of data and materials}

The data-sets used and analysed during the current study are available from the corresponding author on reasonable request.

\bibliography{sn-bibliography}

\end{document}